# Irreversible transformation of ferromagnetic ordered stripe domains in single-shot IR pump - resonant X-ray scattering probe experiments

Nicolas Bergeard[1], Stefan Schaffert[2], Víctor López-Flores[1,3], Nicolas Jaouen[3], Jan Geilhufe[4], Christian M. Günther[2], Michael Schneider[2], Catherine Graves[5,6], Tianhan Wang[5,7], Benny Wu[5,6], Andreas Scherz[5], Cédric Baumier[3,8,9], Renaud Delaunay[8,9], Franck Fortuna[10], Marina Tortarolo[8,9], Bharati Tudu[8,9], Oleg Krupin[11], Michael P. Minitti[11], Joe Robinson[11], William F. Schlotter[11], Joshua J. Turner[11], Jan Lüning[8,9,3], Stefan Eisebitt[2,4], and Christine Boeglin[1]*

1- Institut de Physique et de Chimie des Matériaux de Strasbourg, UMR7504, CNRS et Université de Strasbourg, 23, rue du Loess, 67034 Strasbourg, France.
2- Institut für Optik und Atomare Physik, Technische Universität Berlin, Straße des 17. Juni 135, 10623 Berlin, Germany
3- Synchrotron SOLEIL, L'Orme des Merisiers, Saint-Aubin, BP 48, 91192 Gif-sur-Yvette Cedex, France
4- Helmholtz-Zentrum Berlin für Materialien und Energie GmbH, Hahn-Meitner-Platz 1, 14109 Berlin, Germany
5- Stanford Institute for Materials & Energy Science (SIMES), SLAC National Accelerator Laboratory, 2575 Sand Hill Road, Menlo Park, California, 94025, USA
6- Department of Applied Physics, Stanford University, Stanford, California 94035, USA
7- Department of Materials Science and Engineering, Stanford University, Stanford, California 94035, USA
8- Sorbonne Universités, UPMC Université Paris 06, UMR 7614, LCPMR, 75005 Paris, France
9- CNRS, UMR 7614, LCPMR, 75005 Paris, France
10- Centre de Spectrométrie Nucléaire et de Spectrométrie de Masse, CNRS/IN2P3, Université Paris-Sud, UMR 8609, 91405 Orsay, France
11- Linac Coherent Light Source, SLAC National Accelerator Laboratory, Menlo Park, California 94025, USA

**Abstract**

The evolution of a magnetic domain structure upon excitation by an intense, femtosecond Infra-Red (IR) laser pulse has been investigated using single-shot based time-resolved resonant X-ray scattering at the X-ray Free Electron laser LCLS. A well-ordered stripe domain pattern as present in a thin CoPd alloy film has been used as prototype magnetic domain structure for this study. The fluence of the IR laser pump pulse was sufficient to lead to an almost complete quenching of the magnetization within the ultrafast demagnetization process taking place within the first few hundreds of femtoseconds following the IR laser pump pulse excitation. On longer time scales this excitation gave rise to subsequent irreversible transformations of the magnetic domain structure. Under our specific experimental conditions, it took about 2 nanoseconds before the magnetization started to recover. After about 5 nanoseconds the previously ordered stripe domain structure had evolved into a disordered labyrinth domain structure. Surprisingly, we observe after about 7 nanoseconds the occurrence of a partially ordered stripe domain structure reoriented into a novel direction. It is this domain structure in which the sample's magnetization stabilizes as revealed by scattering patterns recorded long after the initial pump-probe cycle. Using micro-magnetic simulations we can explain this observation based on changes of the magnetic anisotropy going along with heat dissipation in the film.

# I. INTRODUCTION

Ultrafast magnetization dynamics excited by ultra-short laser pulses have attracted tremendous interest in the past years [1, 2, 3] motivated by the scientifically quest to understand the underlying mechanisms in different material systems as well as the importance of potential applications such as in magnetic recording media. One of the most challenging aspects studied in recent years in this field is the dynamic response of magnetic domains to such excitations. Most of the experiments use the magneto-optical Kerr effect to probe the magnetic state dynamically [4, 5]. More recently, time-resolved experiments exploiting short wavelengths in the XUV and soft X-ray range have enabled element-specific investigations. One of the sources enabling such experiments is a synchrotron radiation storage ring in specific operation modes generating shorter pulses as usual, such as "low alpha" and "femto-slicing" [6-8]. Element-specific experiments at these sources have already brought new insight into the intensely debated subject exploiting the X-ray magnetic circular dichroism (XMCD) effect in a transmission geometry with 130 fs time resolution [6, 7, 9-12]. More recently, XUV sources from high-harmonic laser systems have been used to perform repetitive pump - probe experiments for time resolved spectroscopy [13, 14] and magnetic domain scattering experiments [15] offering a time resolution of down to 50 fs. All of the above mentioned sources, however, suffer from comparatively low photon flux and are therefore limited to repetitive pump-probe experiments, building up signal via averaging over millions of repetitions for the basic experiment.

With non-local effects due to electron (and hence possibly spin) transport during ultrafast optical demagnetization being reported and debated, the spatial arrangement in different materials and within magnetically ordered regions becomes important [10, 13, 15-20]. Consequently, the combination of temporal with spatial and/or elemental resolution has become relevant [13, 19, 20]. A particular but important case of magnetic order is represented by quasi-two dimensional domain networks. Domains present in magnetic thin films with perpendicular anisotropy fall into this category and have been studied in many static and dynamic XMCD-based experiments in the past [19-24]. In fact, this class of samples constitutes a widely used model system due to (i) the tunability of their material properties via the multilayer composition; (ii) the possibility to modify the magnetic domain pattern via the magnetic history, e.g., preparing a disordered labyrinth or ordered stripe domain phase, (iii) the flexibility to introduce different magnetic elements and magnetic coupling schemes [25, 26]. A further advantage offered by such films is the strong magnetic contrast accessible in scattering experiments due to the pronounced XMCD effect of the absorption resonance of these elements in the soft X-ray photon energy range [27-29].

Given the currently widespread use of such films for time resolved experiments on ultrafast time scales, which are mostly integrating multi-shot experiments, a potential change of the spatial arrangement of the magnetic domain structure is a relevant concern [30, 31]. For this reason, repetitive pump-probe experiments are limited to the regime of weak pump fluence, i.e. well below the threshold for sample modifications. Single-shot experiments, on the other hand, allow exploring higher pump fluence values and thus to explore unique pathways for rearrangement of a magnetic domain structure, i.e., the occurrence of irreversible magnetization changes which cannot be investigated in experiments relying on repetitive pump-probe cycles to accumulate signal statistics. We note that although experiments following the irreversible modification of a magnetic domain structure following a nearly complete quenching of the magnetization have not been reported so far, the underlying physical processes have been addressed already theoretically [32].

Here we report on such an experiment where a single pump pulse with a fluence of 65 mJ/cm$^2$, induces an almost complete quenching of the magnetization within the first picosecond. The subsequent dynamics occurring on the nanosecond time scale lead to a loss of the magnetic stripe domain phase order. Our results provide the important time scales and the following time-dependent quantitative values: average domain size, domain width distribution, correlation length, the degree of magnetic domain order and the direction of preferential domain orientation. We note that since such irreversible transformations could exist for lower pump fluencies and that they occur on the nanosecond time scales they may remain unobserved in repetitive pump-probe experiments and thus affect the interpretation.

## II. EXPERIMENTAL

To characterize local magnetization magnitude and magnetic order we employed resonant small-angle X-ray scattering (SAXS) at the Co L$_3$ resonance, which exhibits a strong X-ray magnetic circular dichroism (XMCD) [29, 31]. Magnetic domain structures with a characteristic periodicity as those present in our CoPd alloys have been used in several resonant magnetic scattering experiments before [15, 19, 31, 33, 34]. In such experiments SAXS provides insight in static domain sizes and magnetic domain order [25, 29] or follows the dynamics during the ultrafast demagnetization process [15, 19]. We extend this approach here to investigate the recovery of the magnetic order after an initial nearly complete magnetization quenching induced by a single intense femtosecond IR pump pulse. The samples consist of 50 nm thin Co$_{60}$Pd$_{40}$ alloy films sputter-deposited on Si$_3$N$_4$ square membranes of 50 μm in size and 100 nm in thickness. With Si chips holding about 60 such membranes, all investigated films of this study have been deposited under identical conditions. A capping layer of a 3 nm Pd film has been used to protect the ferromagnetic film against oxidation. The magnetic properties of each film were characterized using magneto-optical Kerr effect measurements (MOKE), static resonant magnetic SAXS and magnetic force microscopy (MFM). This showed that the films do exhibit at room temperature the expected strong perpendicular magnetic anisotropy (PMA) [35] stabilizing a domain network of alternating up and down oriented spins in adjacent domains. An in-plane demagnetization procedure using a decreasing oscillating magnetic field has been employed in order to realize a meta-stable magnetic domain structure of strongly aligned stripe-domains (see MFM image in Figure 1a). The observed momentum transfer in the scattering pattern reveals an average magnetic domain width of 71 ± 5 nm (Figure 1a), which is in line with the local areas observed in the MFM images. For each thin film the following sequence of X-ray scattering patterns has been recorded: (i) The initial state of the film's magnetic domain structure was characterized by recording a scattering pattern with a single X-ray pulse only; (ii) the single shot IR-pump SAXS-probe experiment was performed for a specific delay; (iii) the final state of the rearrangement of the magnetic domain structure was characterized with a scattering pattern recorded with a single X-ray pulse, seconds after the previous pump-probe cycle. We refer in the following to the first and last step as pre- and post-characterization. We note that using LCLS's gas attenuator, the X-ray pulse intensity was adjusted such that sufficient intensity for a single shot scattering experiment was provided, but no X-ray induced sample changes were observed. The feasibility of reaching such a regime has been shown previously, e.g., in Ref. [31]. A sketch of the experimental geometry is given in Fig. 1.

In the presence of a well-aligned magnetic stripe domain structure the X-ray diffraction pattern exhibits two intense spots, which correspond to the plus and minus first diffraction order of the grating-like magnetic domain structure (Fig.1b). A labyrinth domain structure, on the other hand, will give rise to a ring-like scattering intensity distribution due to the isotropic orientation of domains within the sample plane [19]. A superposition of two Bragg spots and a ring structure is consequently obtained for a magnetic domain structure exhibiting a partial/preferential orientation of the magnetic domains. More generally, we note that the degree of domain order is reflected in the angular anisotropy of the diffraction pattern. To extract the degree of domain alignment from the scattering pattern we use azimuthal line cuts of the scattering intensity integrated over the q-range of the diffraction peak region (q-range indicated in gray in Fig. 1c). This yields the azimuthal intensity distribution of the scattering intensity shown in Fig. 1d, which will be used in the following discussion.

The experiment was carried out in the RCI chamber at the SXR beamline [36] of the Linac Coherent Light Source (LCLS) X-ray free-electron laser, which is located at the SLAC National Accelerator Laboratory. The SXR monochromator was used to obtain monochromatic X-ray pulses of 778 eV, matching the magnetically dichroic Co $L_3$ absorption resonance. The X-ray pulse length as estimated from the electron bunch length was 300 fs. A single X-ray pulse has been delivered on request, either with or without a preceding IR pump pulse. The X-ray pulses were focused with a Kirkpatrick-Baez mirror pair to an elliptical spot size of 10 x 15 µm². Using the facility's gas attenuator the pulse intensity was adjusted such that sufficient intensity for a single shot scattering experiment was provided, but no X-ray induced sample changes were observed. The feasibility of reaching such a regime has been shown previously, e.g., in Ref. [31]. We remark that despite improving capabilities developed at the SXR instrument, a precise measurement of the incident photon flux remains still difficult. This is in particular the case in experiments like ours, where the sample itself presents an additional spatial aperture and spatial beam fluctuations result in additional, not detected intensity fluctuations. From the recorded scattering patterns we can retrieve as an estimate for the incident photon flux a value of $10^5$ - $10^7$ photons/ µm². This corresponds to 0.1 – 0.001 mJ/cm², which is well-below the threshold for X-ray induced sample modifications for our type of samples [31].

The scattering patterns were detected with a soft X-ray CCD camera (MTE, Princeton Instruments, 2048 x 2048 pixels of 13.5 x 13.5 µm²), which was positioned 520 mm downstream of the sample. A 400 nm thin Aluminium filter was used to render the camera light tight, in particular, against the IR photons from the pump pulses. The chosen geometry and employed X-ray wavelength determine the maximum detectable scattering vector to be 103 µm$^{-1}$, which corresponds to a wavelength of spatial variations of 61 nm.

The intense direct transmitted beam was blocked by a tungsten-carbide rod, which was attached to a steel sheet. As can be seen in the scattering patterns shown in Fig. 2a-c, the beam stop was oriented such that it does not cover the scattering peaks originating from the original aligned stripe phase. The intense, horizontal scattering lines in the scattering pattern are due to charge scattering from the membrane edges and indicate that the horizontal wings of the X-ray beam overfill the Si3N4 membranes. We note that the intensity of this charge scattering depends strongly on the relative positioning of the membrane and the incident X-ray pulse (and also on the microstructure of the edges of the membranes). For this reason, it cannot be used to estimate the intensity of the incident X-ray pulses.

Pump pulses were provided by the LCLS Ti:Sapphire IR laser ($\lambda$=800 nm) with a pulse energy of 86 µJ and a duration of about 50 fs, with a default setting of linear, horizontal polarization. Spatial overlap of the X-ray and IR laser beam was achieved by focusing both beams on a Ce:YAG screen. For precise determination of the temporal overlap a cross-correlator measurement observing the IR reflectivity of a tilted silicon-nitride plate as a function of the delay of an intense X-ray illumination was used [37]. The transverse laser beam profile was circular and close to a Gaussian shape with a measured FWHM of 340 µm, which yields rather homogeneous IR fluence over the whole sample membrane in the chosen normal incidence geometry. The resulting incident IR laser fluence on the open membrane was determined from the Gaussian profile to have an average value of 68±7 mJ/cm². The first interface vacuum - Pd-capping has a reflectivity of 76% for 800 nm (1.55 eV)

(bulk values from Ref. [38]) leading to a transmitted IR intensity of about 16±2 mJ/cm² into the CoPd layer. In our discussion below we assume constant refractive index and specific heat values for all constituents at all times.

## III. RESULTS

In absence of a proper measurement of the intensity of the X-ray probe pulse, we have used for visualization of the scattering pattern the charge scattering from the membrane edges to obtain an approximate, relative normalization of the scattering patterns. Prior to this all images have been background corrected by subtracting a reference dark image. We remark that this normalization has no influence on the analysis and interpretation of the data, since our discussion focuses on changes in the spatial distribution of the observed scattering intensity. For this we extract from the images the radial $I(q)$ and azimuthal $I(\Phi)$ scattering intensity distribution (figure 1c and d). We note that for the calculation of the radial distribution $I(q)$ the regions of the beamstop and the charge scattering have been masked out. A mask with point-symmetry to the q=0 origin of the scattering pattern was used to derive the azimuthal distribution $I(\Phi)$. Figure 1c shows that the radial intensity distribution $I(q)$ exhibits a peak at about 44 µm$^{-1}$, which originates from the presence of the magnetic domains with a characteristic width D. This is related to the momentum transfer $q_{peak}$ by $2D = 2\pi / q_{peak}$ [29], i.e., we find an average domain width of 71 nm in the stripe phase. The variation of the domain width present in the sample is characterized by the width of the peak $I(q)$, which we determine to correspond to 6 nm (FWHM) in real space. We note that this is in perfect agreement with the static pre-characterization of our samples.

The azimuthal intensity distribution $I(\Phi)$ yields a characterization of the ordering and the correlation length of the magnetic domains. In order to quantify the degree of ordering, we define the orientation anisotropy $\vec{K}=(I_{max}(\Phi)-I_{min}(\Phi))/(I_{max}(\Phi)+I_{min}(\Phi))$. We then have $\vec{K}=1$ for perfectly aligned stripe domains and $\vec{K}=0$ for an isotropic labyrinth domain structure. Typical values for our films prior to laser excitation are $\vec{K} = 0.98$ with an in-plane correlation length $2\pi/\Delta q$ ~900 nm [29].

In figure 2a-c, we show the scattering patterns for time delays of 2.5, 5.5 and 7 ns, respectively. At 2.5 ns (Fig. 2a) one observes a rather weak magnetic scattering, which indicates that the magnetization is still strongly quenched. The spatial distribution of the scattering intensity, on the other hand, resembles with two spots on the vertical axis the one of the initial state. For other directions, but at corresponding momentum transfer value, only a weak additional contribution is noticed. The orientation anisotropy in this configuration is measured to be $\vec{K}=0.68\pm0.15$. One further notices, close to the beam stop in the vertical direction, two pronounced streaks at low q at the azimuthal angle corresponding to the maximum intensity at higher q. Since these are absent in the post-characterization scattering pattern (Fig. 2c) and also different from the pre-characterization pattern (Fig. 1b), we have to conclude that they are not due to charge scattering from the upper and lower membrane borders, but must be of magnetic origin. Their appearance is however, not of direct relevance to the reorientation of the magnetic domain structure presented here, and will thus be discussed elsewhere. At a delay of 5.5 ns (Fig. 2b) the magnetic scattering intensity has mostly recovered. One notes that the spatial distribution exhibits a scattering ring with slightly enhanced intensities at the angles found in the pre-characterization pattern (at 0° and 180°, see Fig. 1b for geometry definition). This suggests that the magnetic domain structure at this point in time, exhibits a partially disordered stripe domain phase, i.e., at a delay of 5.5 ns, we have obtained a snapshot of the evolution from a well-aligned stripe phase towards a labyrinth phase.

The numerical analysis of the time resolved images (Fig. 2a,b) shows that the initial value of the orientation anisotropy $\vec{K} = 0.980\pm0.002$ has reduced to $\vec{K} = 0.24\pm0.18$ at the delay of t = 5.5 ns. The radial intensity distribution $I(q)$ (not shown) features also a broader peak at 47.5 µm$^{-1}$ characteristic for a larger width distribution ($2\pi/\Delta q$~300 nm) and a smaller average domain width of 66 nm (compared with 71nm at t<$t_0$). This transient state is attributed to the residual thermal energy still present at this time in the film. The domain walls are not directly observed in this experiment, but can be assumed to be broader than in the unpumped state, which will be seen in the micromagnetic calculations discussed below. Note that domain wall broadening after IR-pumping has also been observed on a sub-ps time scale in the context of spin-dependent transport of laser excited electrons in magnetic domain patterns recently [19].

Even more pronounced changes are observed in the scattering images at delays of t ≥ 7 ns. At t = 7ns (Fig. 2c) we observe an angular redistribution of the intensity along the previous ring, as highlighted in the azimuthal plot in figure 2d (transition from the blue to the orange line). The initial Bragg spots at Φ=0° (180°) have disappeared, and new ones occurred at Φ=-60° (120°). This qualitative change in domain configuration during the magnetization recovery is delayed by more than 1 ns from the transient domain configuration observed at t = 5.5ns. The numerical analysis of the time resolved images (Fig. 2c) shows that the orientation anisotropy $\vec{K}$ = 0.30±0.15 is similar to the value extracted at t = 5.5ns whereas the direction is now rotated by -60 degrees.

The post-characterization SAXS pattern with their respective azimuthal distributions (Fig. 3a,b) show that this new magnetic domain distribution corresponds to the final state at room temperature. Note that the large intensity variations in figure 3a are due to the shot to shot X-ray intensity fluctuations, precluding an analysis of the relative scattering intensities as a function of pump-probe delay. The average orientation anisotropy extracted from these static images is $\vec{K}$ =0.25 ± 0.15, indicating that the orientation anisotropy and the new angular redistribution of the domains are both "frozen" at values similar to those observed at t= 7 ns. We can conclude from the post-characterization images that the single intense IR pump pulse has irreversibly transformed the initial stripe order to a new magnetic order.

To learn more about the final magnetic domain structure, we have used magnetic force microscopy (MFM) to image the final state of the films, which transient scattering pattern is shown in Fig. 2a-c. In figure 3c we show the MFM image (field of view of 20 x 20 µm²) recorded on the trace left by the laser spot, i.e., within the area pumped by the single IR pulse. One of the membrane edges served as the common reference for absolute azimuthal orientation in the MFM and scattering measurements. To extract the preferential domain orientation and the order parameter $\vec{K}$ we have performed a 2-dimensional Fourier Transformation of the image. The obtained data, shown by the blue points in Fig. 3d, indicate a preferential orientation with an angle of -60 degrees, confirming the orientation anisotropy of the final Bragg spots observed by SAXS.

The observed preferential domain reorientation occurring around 7ns calls for a mechanism defining this specific in-plane axis. The IR pump laser polarization can be ruled out as a symmetry-breaking factor since we have checked the influence of IR polarization using linearly as well as circularly polarized IR pump pulses without observing different experimental outcomes. Neither was the membrane size or edge orientation relative to the irradiated spot correlated to the observed reorientation direction.

The presence of a magnetic field, on the other hand, would be an obvious factor to break the symmetry in the experiment and influence domain orientation. While no magnetic field was applied deliberately, the experiment was carried out in a vacuum chamber not shielded against magnetic fields and significant amount of steel surrounds the experimental end station in the walls of the underground building of the LCLS experimental hall. We also note that the Earth's magnetic field has a nearly vertical component of about 50 µT at the site of LCLS [40]. We therefore measured the magnetic field at the sample position using a Gauss meter and found indeed a non-vanishing in-plane magnetic field component. Within this plane, we find the largest value of 120 ± 20 µT along the in-plane direction Φ = +30 degree (as defined in figure 1), whereas the orthogonal in-plane and the out-of-plane directions exhibit smaller values of 50 ± 20 µT. We thus conclude that a symmetry breaking magnetic field has been present during our experiment and that the direction of it's strongest component is indeed along the direction of the preferential reorientation of the magnetic domains.

To understand how a small external magnetic field might lead to the realignment of the magnetic stripe domains, we consider the sample at the recovery time of 5.5 ns when long-range magnetic order starts to be restored (Fig. 2b). In CoPd alloys, the magnetic anisotropy is reduced when the temperature is increased towards the Curie temperature [7]. The high temperature state results also in large domain walls as evidenced by the radial intensity distribution I(q) at t = 5.5 ns. Therefore, any external magnetic field aligned in the plane of the film would act more efficiently on the film's magnetization during this transient high temperature state and might thus align the stripes preferentially along the external field direction. To support this idea, we have performed micromagnetic simulations using the OOMMF software [41], which determines the magnetic configuration as a function of time by solving the time-dependent Landau Lifschitz Gilbert (LLG) equation [42]. The magnetic parameters for the CoPd alloy have been taken from literature [22, 43, 44].

In figure 4a and b we show the simulated magnetic configurations for a 50 nm thick CoPd alloy film obtained for two different values of the magnetic anisotropy constants defining the system at room temperature and at an elevated temperature, $K_m$=5e6 J/m$^3$ and $K_m$=1.6e5 J/m$^3$, respectively. Comparing the two simulations we identify the effect of temperature on the domain configurations. For both images (Fig. 4a and b), we set the saturation magnetization to $M_S$ = 8.7e5 A/m and the exchange stiffness to A = 9e-12 J/m. In figure 4c, we show the magnetization profile between an up and down domain. As expected, we obtain wider domain walls at an elevated temperature (low anisotropy constant), which goes along with an increase of in-plane spins relative to the domain structure at lower temperatures.

In order to reproduce the evolution of the domain reorientation observed between 5 ns and 7 ns we simulated an isotropic labyrinth domain pattern at an elevated temperature ($K_m$=1.6e5 J/m$^3$). The obtained magnetic domain structure for certain points in time are shown in Fig. 5. In the case of the simulations shown in Fig. 5a-c, we introduced an in-plane anisotropy by applying a magnetic field of 100 µT along a given in-plane direction, while no external magnetic field is present in the case of the simulations shown in Fig. 5d-f. In order to avoid boundary effects, the evaluation of the micromagnetic simulations are limited to the central part indicated by the red circles in Figure 5. Our simulations show that the combination of the lowered magnetic anisotropy constant in conjunction with the in-plane magnetic field is sufficient to give rise to an evolution of the magnetic domain structure, which leads to a preferential orientation of the magnetic domains parallel to the externally applied magnetic field (see Figure 5c). We can thus conclude that the micromagnetic simulation supports the proposed mechanism.

Our simulations shown in figure 5c indicate that maintaining the film artificially at high temperature (low anisotropy) up t= 50 ns would further enhance the transformation and alignment of the domains. In the experiment, on the other hand, sample cooling back towards room temperature proceeds at the same time and the anisotropy thus increases with time so that the evolution of the domains will be suddenly locked. The comparison with the post-characterization images suggests that this occurs at about 10 ns. This delay time can further be confirmed by the fact that the orientation anisotropy reaches the final value at about 10 ns (Figure 6). While, in a ferromagnetic material, it is generally not surprising that magnetization can be easier aligned by an external field at elevated temperatures towards the Curie temperature, our simulations show that such a process can indeed explain our observations in the CoPd system under study.

## IV. CONCLUSION

We have performed single-shot IR-pump - X-ray probe experiments on meta-stable aligned stripe domain structures in CoPd alloy films. Our results show the dynamics of the evolution of the magnetic domain structure during quenching and recovery. The single, intense laser excitation triggering the well-studied ultrafast demagnetization leads to an almost complete quenching of the magnetization on a sub-picosecond time scale. During the subsequent magnetization recovery, irreversible transformations of the magnetic domain structure occur. This proceeds through a sequence of new magnetic domain configurations on the time scale of a few nanoseconds after the excitation. The initial, well-aligned stripe domain structure evolves towards a disordered labyrinth domain phase. Delayed by ~1 ns, a second transformation, characterized by a new angular distribution of the domains, is observed after t=7 ns. Micromagnetic simulations attribute this recovery process to the low anisotropy of the film during the recovery in the presence of an external magnetic field. We thus show that a simple manipulation tool (IR laser pulse) can be used to induce a new magnetic order in magnetic domains by orienting the domains with respect to a weak external magnetic field. This transformation of the magnetic domains takes place on the time scale of a few nanoseconds. Experiments with infrared or X-ray excitation of magnetic domains integrating over numerous pump-probe cycles at elevated pump fluences [31] have to take such potential domain configuration changes into account.

## ACKNOWLEDGMENTS

We are indebted to J.Y. Chauleau for help and support for the micromagnetic simulations and to A. Boeglin for discussions and correction of the manuscript. The authors are grateful for financial support received from the following agencies: the German Federal Ministry of Education and Research under contract BMBF-05K10KTB/FSP-301 ("MPScatt"); the PEPS SASLEX of the CNRS (France); the Université de Strasbourg and the French "Agence Nationale de la Recherche" via the project EQUIPEX UNION: # ANR-10-EQPX-52", and the Ministry of Education of Spain (Programa Nacional de Movilidad de Recursos Humanos del Plan Nacional de I-D+i 2008-2011). Portions of this research were carried out on the SXR Instrument at the Linac Coherent Light Source (LCLS), a division of SLAC National Accelerator Laboratory and an Office of Science user facility operated by Stanford University for the U.S. Department of Energy. The SXR Instrument is funded by a consortium whose membership includes the LCLS, Stanford University through the Stanford Institute for Materials Energy Sciences (SIMES), Lawrence Berkeley National Laboratory (LBNL), University of Hamburg through the BMBF priority program FSP 301, and the Center for Free Electron Laser Science (CFEL).

(*) corresponding author: christine.boeglin@ipcms.unistra.fr

Figures and captions:

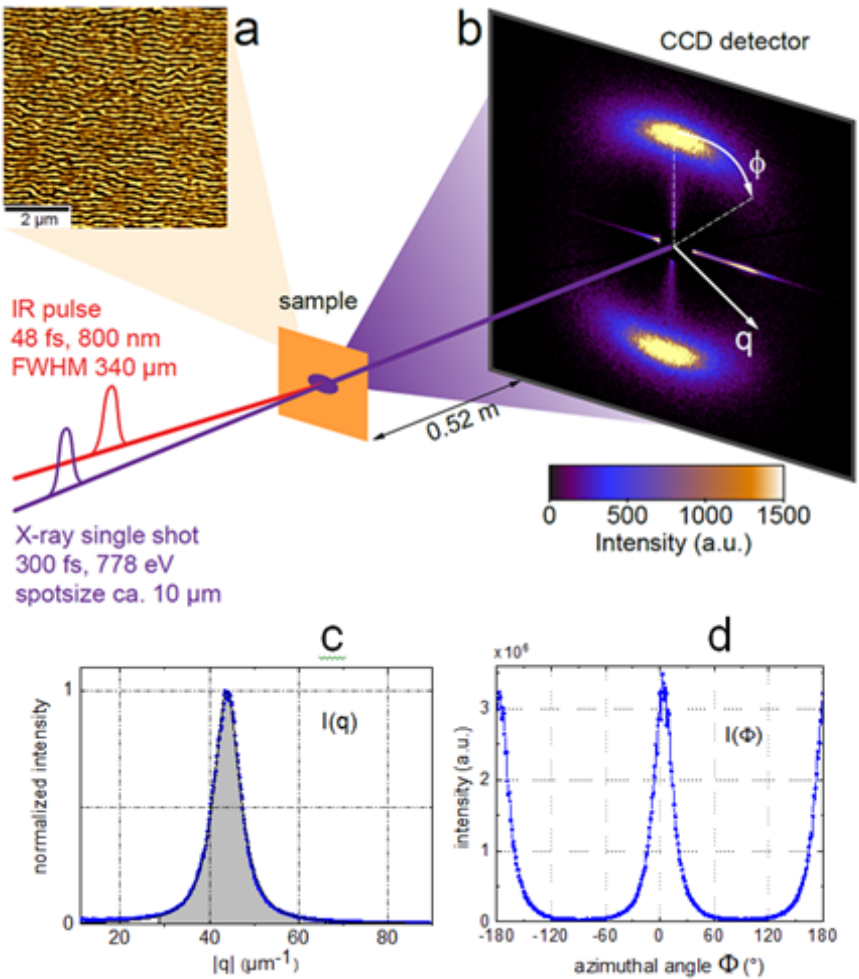

Figure 1: Sketch of the experimental setup. Each CoPd alloy membrane is illuminated with a single-shot IR pump and a single X-ray probe pulse. A post-characterization is performed with a single X-ray probe pulse at a delay of t = 1 minute. The small-angle X-ray diffraction pattern of the magnetic domains is recorded on a CCD detector. a) Magnetic force microscopy (MFM) image of the prepared stripe domains with an average domain size of 71 nm. b) Corresponding small-angle X-ray scattering pattern with scattering vector q and azimuthal angle Φ. c) The radial intensity distribution I(q) defines the degree of homogeneity in the domain width. d) The azimuthal angle intensity distribution I(Φ) defines the degree of alignment of the domains in the plane of the film. The intensity distribution I(Φ) defines the quantitative orientation anisotropy constant $\vec{K}$ which is 0.98 in the prepared state before excitation by the pump laser pulse (b).

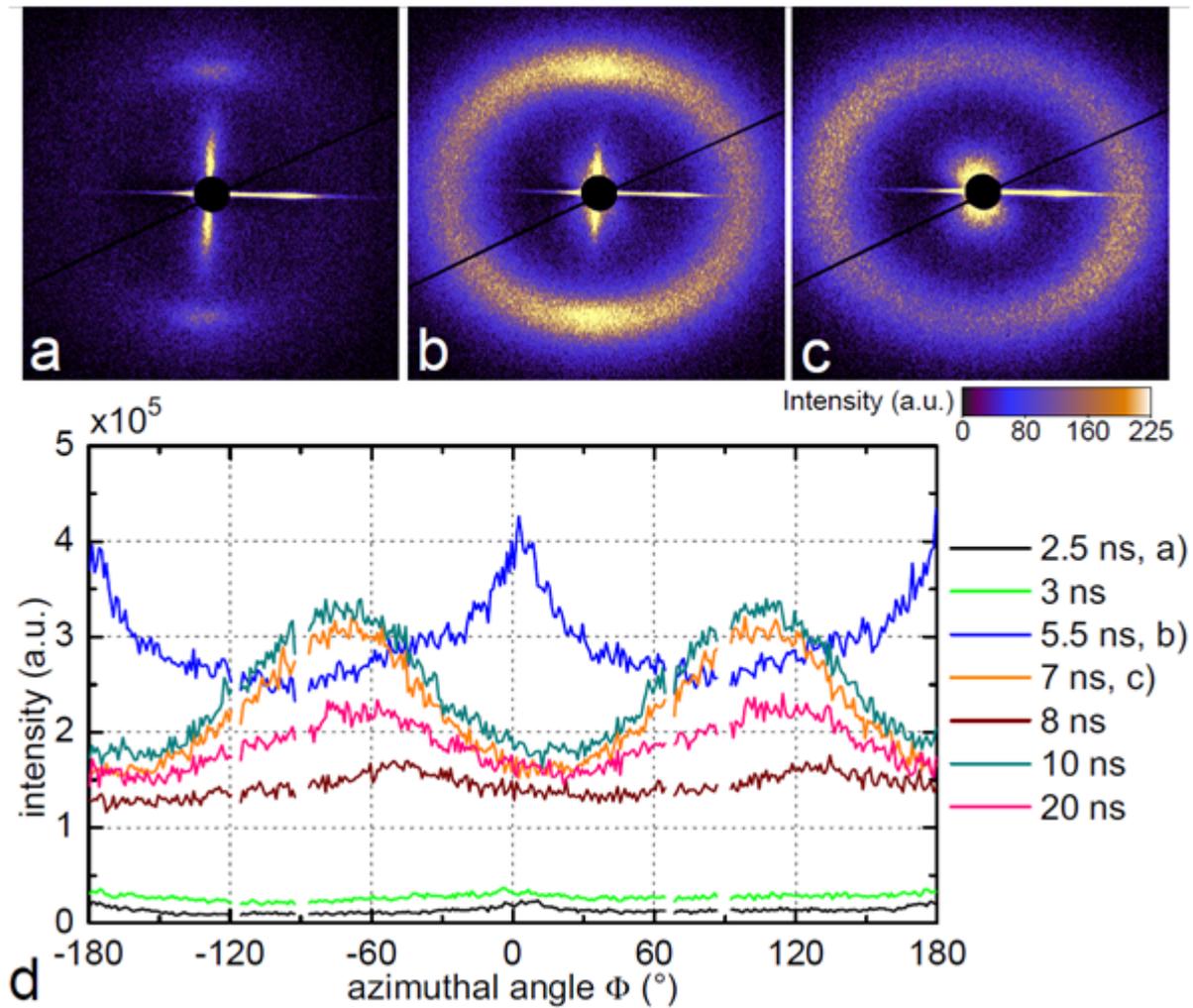

Figure 2: (a, b, c) IR pump - X-ray probe diffraction patterns for different delay times: t= 2.5 ns, 5.5 ns and 7 ns. a) The almost complete loss of the magnetic scattering intensity (1ps < t <3ns) is observed after an intense single IR laser pulse. b) Partial recovery of the magnetic scattering intensity at t=5.5 ns showing that disordered labyrinth domains are formed and that partially ordered stripes are recovered aligned along the *initial* direction. c) New partially ordered stripes are observed at t = 7 ns with a 60° rotation of the stripe alignment direction. d) Azimuthal intensity distributions I(Φ) for different single shot pump-probe delays. A new preferential orientation direction is observed at t ≥7 ns as evident from the by superposition of a scattering ring and two Bragg spots. The new magnetic phase order is characterized by an orientation anisotropy of $\vec{K}$ = 0.25 ±0.15.

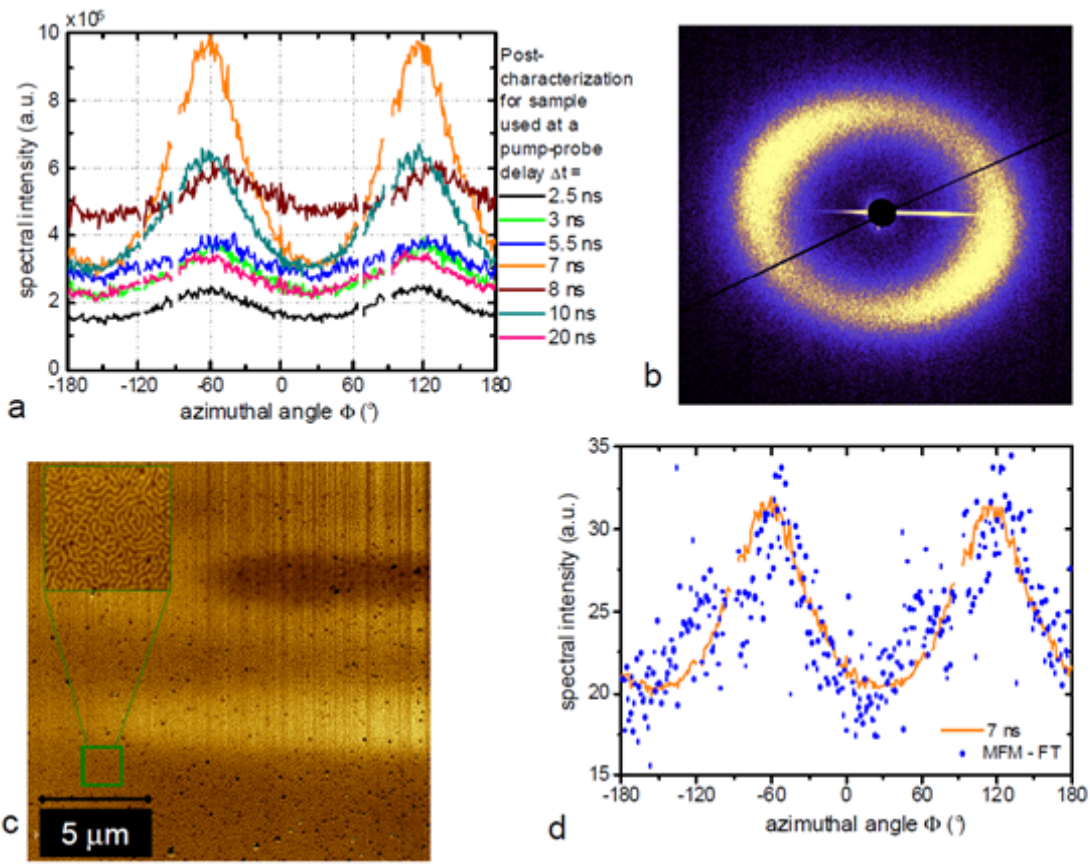

Figure 3: a) Post-characterization azimuthal angle intensity distribution $I(\Phi)$ for all investigated films at room temperature. The new angular distribution (with the preferred axis rotated by -60 degree) is found independent of the pump-probe delay used during the experiment. b) Scattering pattern of the final state observed for the film used in the pump-probe experiment with a pump-probe delay of t =7 ns. c) Magnetic force microscopy (MFM) image at room temperature recorded in a section of the area exposed to the IR pulse of the aforementioned sample. d) The Fourier transformation of the MFM image (blue markers) reveals an anisotropic, preferential orientation of the magnetic domain structure, which is in perfect agreement with the one characterized by the azimuthal line cut of the post-characterization image of the same sample (orange line)..

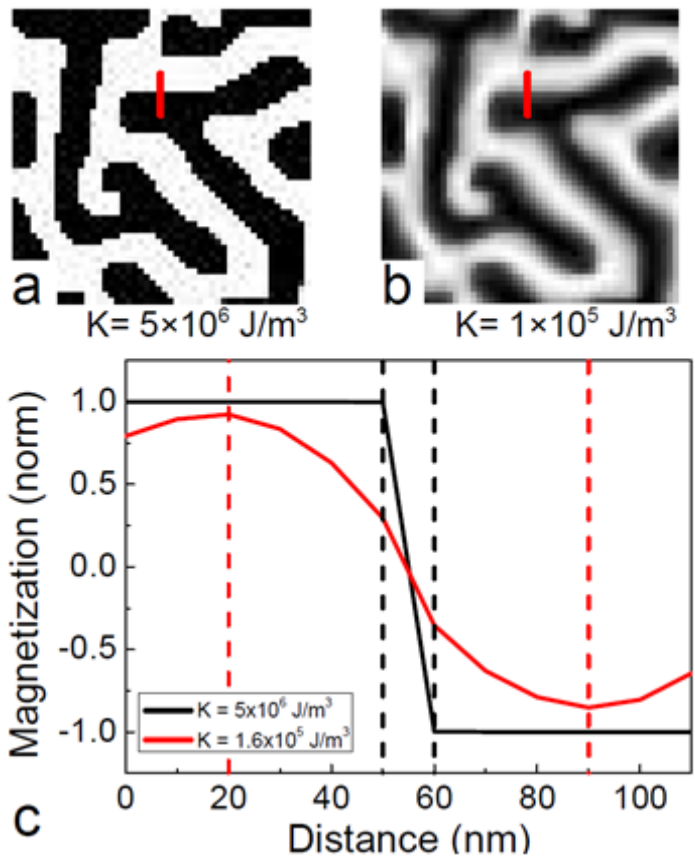

Figure 4: Evolution of the domain wall profiles with the temperature. a) Worm domains at 300 K (anisotropy constant K = 5e6 J/m$^3$) and b) worm domains at elevated temperature (K = 1.6e5 J/m$^3$). c) Domain wall profiles extracted along the red line in images (a) and (b). The black profile line is extracted from (a) and the red line from image (b). The magnetization amplitudes are normalized to those in the image (a). The larger extension of the wall size is observed at elevated temperature. The image size is 0.5 x 0.5 µm². The spatial resolution of the simulation is 5 nm.

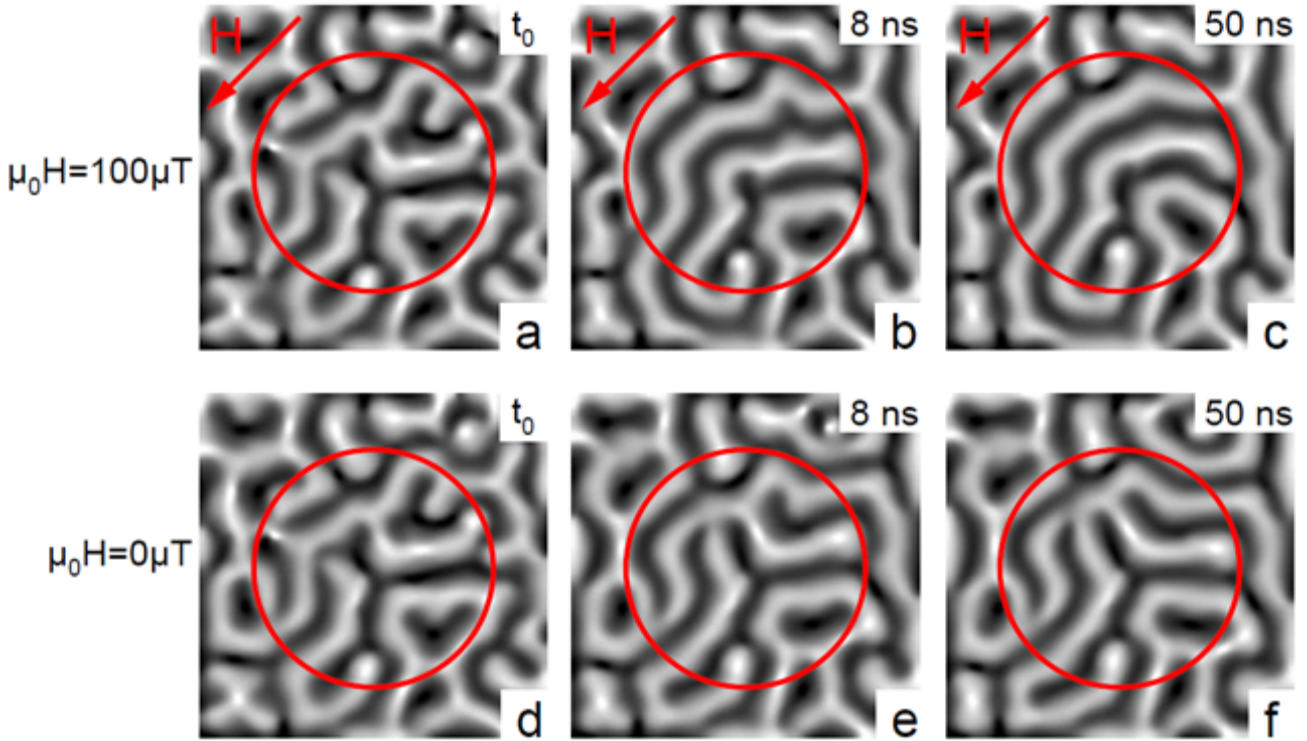

Figure 5: Micromagnetic simulations investigating domain dynamics following the Landau-Gilbert time-dependent evolution. a) The starting configuration analogue to the recovery of the magnetization at t= 5.5 ns (see figure 2b). b)-c) Simulated temporal evolution in the presence of an in-plane anisotropy field of 100 µT (red arrows). d)-f) Analogous simulation of the temporal evolution of the labyrinth domains obtained without in-plane anisotropy field. The size of the images is 1 x 1 µm² and the resolution is 5 nm. The red circles limit the centre of the calculated images to avoid boundary effects. Inside the circle the magnetic domains are found to be aligned preferentially parallel to the external magnetic field (the direction is given by the red arrows).

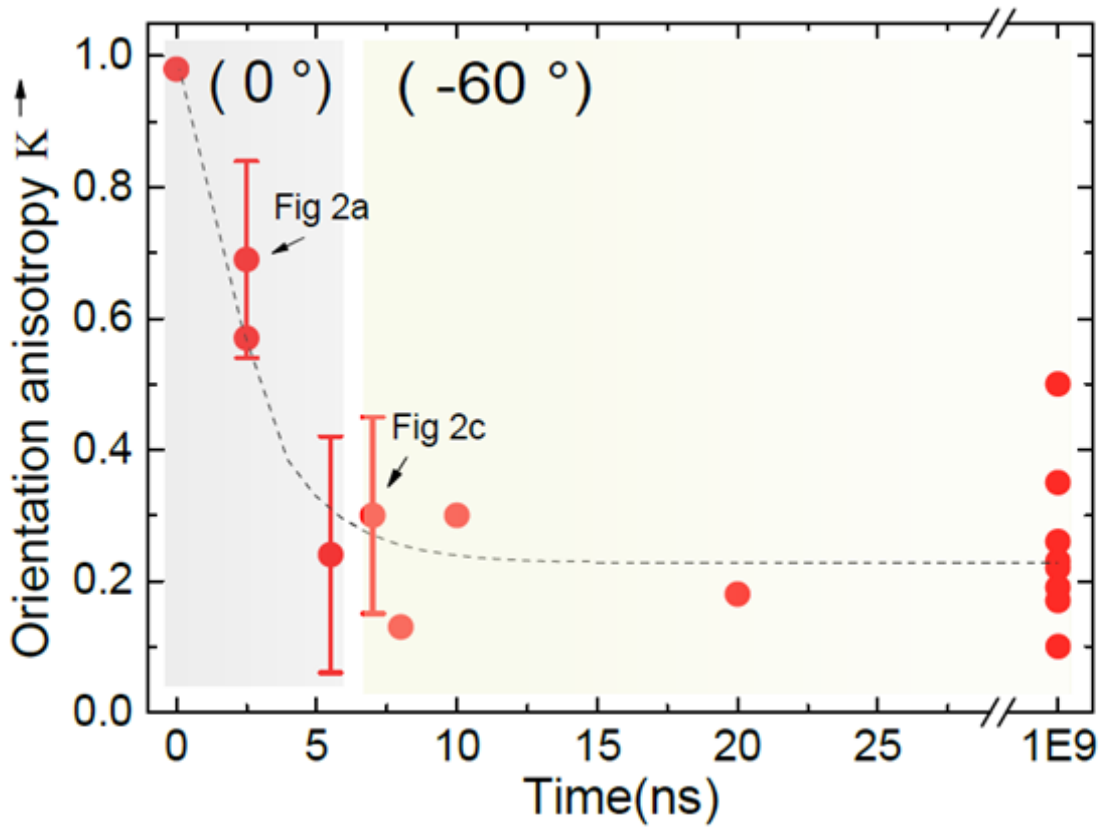

Figure 6: Dynamic of the orientation anisotropy $\vec{K}(t)$ characterizing the dynamics of the partially ordered domain structure. The large dispersion of the data points is related to the lateral instability of the X-ray pulses modifying the thermal recovery and the magnetic anisotropy of the film. The dashed curve is given as a guide for the eyes. The grey and yellow colours highlight the observed angular redistribution of the intensity along the resonant scattering rings from $\Phi = 0°$ to $\Phi = -60°$ corresponding to the preferential reorientation of the domains along the in-plane direction from $\Phi = 0°$ to $\Phi = +30°$. As the time interval during which the film is at low magnetic anisotropy is short (between $t \sim 5$ ns and $t \sim 10$ ns), the orientation anisotropy is only partial. The error bars for $\vec{K}(t)$ are given by the standard deviation of the experimental data points.